\documentclass[aps,preprint,epsfig,rotate]{revtex4}
\begin{document}
\title{On the Hamilton approach for the metric GR}

 \author{Alexei M. Frolov}
 \email[E--mail address: ]{afrolov@uwo.ca}

\affiliation{Department of Applied Mathematics \\
 University of Western Ontario, London, Ontario N6H 5B7, Canada}

\date{\today}

\begin{abstract}

Basic principles of the Hamilton approach developed for the metric General Relativity (Einstein`s GR) are discussed. In particular, we derive the Hamiltonian of the 
metric GR in the explicit form. This Hamiltonian is a quadratic function  of the momenta $\pi^{mn}$ conjugate to the spatial components $g_{mn}$ of the metric 
tensor $g_{\alpha\beta}$. The Hamilton approach is used to analyze some problems of metric GR, including the internal structure of propagating gravitational waves and
quantization of the metric GR. We also derive the Schr\"{o}dinger equation for the free Gravitational field and show that actual gravitational field cannot 
propagate as pure harmonic oscillations, or harmonic gravitational waves. A number of inequalities useful in applications to the metric GR are derived.

\noindent 
PACS number(s): 04.20.Fy and 11.10.Ef

\noindent 
15.09.2015, Preprint-2015-8-1/28 [Field Theory; Gravitational Field(s)], 20 pages. \\
First Added: 2015-09-16 T 00:49:25 UTC \\
DOI: 10.13140/RG.2.1.4912.4322

\end{abstract}

\maketitle
\newpage

\section{Introduction}

In this communication we consider the Hamilton approach to the metric General Relativity (GR), or, in other words, to Einstein's GR. Our goal below is to summarize facts known for 
this approach and its applications to the metric GR. Moreover, based on the Hamiltonian of the gravitational field we re-derive some results obtained in earlier studies. By 
considering the Hamilton approach to the metric GR we try to follow an analogy between field equations for the Maxwell electrodynamics and metric GR \cite{Dir50}, \cite{Dir64}, 
\cite{Fro1}. In general, the approach based on the Hamiltonian(s) derived for the metric gravity has a number of advantages in actual applications. In particular, the Hamiltonian 
approach to the metric GR allows one to perform quantization of the metric gravity \cite{Dir50}, \cite{Dir64}. In this study the Hamilton approach means the method which is based 
on the explicit expression(s) for the Hamiltonian(s) of the free gravitational field.

The first attempt to formulate the Hamilton approach for metric gravity was made in \cite{Pir52}, i.e. almost immediately after publication of the famous paper by Dirac \cite{Dir50}, 
where he formulated his famous `constrained dynamics'. However, the method developed in \cite{Pir52} was obviously incomplete. Furthermore, the paper \cite{Pir52} contains a number of 
principal mistakes, which can be found, e.g., in all secondary constraints derived in \cite{Pir52}. These mistakes have been corrected in 2008 when the fundamental paper by Kiriushcheva, 
Kuzmin et al \cite{Kuzm2008} was published. The paper \cite{Kuzm2008} contains a complete and correct version of the Hamilton approach developed for the metric GR. Note that in 1958 
Dirac issued another paper \cite{Dir58} in which he also developed his Hamilton approach for the metric GR. In fact, the approach formulated in the paper \cite{Dir58} was quite different 
from the approach developed in \cite{Pir52} and \cite{Kuzm2008}. Originally, the source of these differences was not clear. The approach developed in \cite{Dir58} was considered as 
the Dirac's version of the Hamilton approach to the metric GR. Finally, in 2011 \cite{Fro2011} we have found that these two Hamiltonian-based approaches to the metric GR, i.e. approach 
from \cite{Pir52}, \cite{Kuzm2008} and `alternative' approach formulated in \cite{Dir58}, are related to each other by a canonical transformation of dynamical variables. But canonical 
transformations of dynamical variables are allowed transformations in the Hamilton procedure. Here and everywhere below in this study by a `canonical tranformation' of dynamical 
variables we mean transformation which conserves numerical values of the Poisson brackets \cite{Gant} - \cite{Golds}. As follows from here the two different Hamilton approaches 
formulated in  \cite{Pir52}, \cite{Kuzm2008} and in \cite{Dir58} (see also \cite{Fro2011}) are, in fact, the two equivalent appearances (or two faces, for short) of the same Hamilton 
approach. Such an explanation was a clear indication of the correctness of the Dirac procedure \cite{Dir58}. Moreover, the conclusion of \cite{Fro2011} shows that now we can develop a 
large number (even infinite number) of different versions of the Hamilton approach to the metric GR which absolutely agree with each other. Note also that only these Hamilton approaches 
allow one to obtain the same gauge invariance of the metric GR (diffeomorphism) which was produced in the Lagrange approach earlier (see, e.g., \cite{Saman}).  

Our analysis in this study includes a brief discussion of the Lagrange procedure which is applied to the free gravitational field in the metric GR. Then we derive the explicit 
formula(s) for the Hamiltonian of the metric GR and discuss the complete version of Dirac's procedure for the constrained dynamical system which represents the actual gravitational 
field(s) in themetric GR. Based on the explicit formula for the Hamiltonian of the metric GR we briefly investigate some long-standing problems in the metric GR. This includes 
investigation of the internal structure of propagating gravitational waves, correct quantization of the metric GR and a few other problems. 

\section{Lagrange approach}

First of all, let us derive the explicit expression for the Lagrangian of the gravitational field. Note that in early years of the metric GR Hilbert and Einstein successfully derived 
Einstein's equations (see discussion and references in \cite{Hilb} and \cite{Einst}) for the free gravitational field by using the `gravitational' action $S_g$ written in the form 
\cite{LLE}
\begin{equation}
  S_g = -\frac{c^3}{16 \pi k} \int L \sqrt{-g} d\Omega = -\frac{c^3}{16 \pi k} \int R \sqrt{-g} d\Omega -\frac{c^3}{16 \pi k} \int \frac{\partial(\sqrt{-g} w^{\gamma})}{\partial 
  x^{\gamma}} d\Omega \label{eq1}
\end{equation}   
where $k \approx 6.67834 \cdot 10^{-8}$ $cm^{3} g^{-1} sec^{-2}$ is the gravitational constant (in $CGS$ units) and scalar $R = g^{\alpha\beta} R_{\alpha\beta}$ is the scalar 
curvature of the space. Based on this formula for $S_g$, Eq.(\ref{eq1}), we can derive the explicit formulas for the Lagrangian ${\cal L} = L \sqrt{-g}$ and for the unknown functions 
$w^{i}$ from Eq.(\ref{eq1}). Indeed, the known expression for the scalar curvature $R$ (or $R \sqrt{-g}$) is well known (see, e.g., \cite{LLE} and \cite{Carm}). It is written in the 
form
\begin{equation}
  R \sqrt{-g} = \sqrt{-g} g^{\alpha\beta} R_{\alpha\beta} =  \sqrt{-g} \Bigl[ g^{\alpha\beta} \frac{\partial \Gamma^{\gamma}_{\alpha\beta}}{\partial x^{\gamma}}
  - g^{\alpha\beta} \frac{\partial \Gamma^{\gamma}_{\alpha\gamma}}{\partial x^{\beta}} + g^{\alpha\beta} \Gamma^{\gamma}_{\alpha\beta} \Gamma^{\rho}_{\gamma\rho} 
  - g^{\alpha\beta} \Gamma^{\rho}_{\alpha\gamma} \Gamma^{\gamma}_{\beta\rho} \Bigr] \label{eq15}
\end{equation}  
The first term in the right-hand side of this equation we transform in the following manner
\begin{equation}
 \sqrt{-g} g^{\alpha\beta} \frac{\partial \Gamma^{\gamma}_{\alpha\beta}}{\partial x^{\gamma}} = \frac{\partial}{\partial x^{\gamma}} \Bigl(\sqrt{-g} g^{\alpha\beta}
  \Gamma^{\gamma}_{\alpha\beta}\Bigr) - \Gamma^{\gamma}_{\alpha\beta} \frac{\partial (\sqrt{-g} g^{\alpha\beta})}{\partial x^{\gamma}} \label{eq151}
\end{equation} 
Analogously, the second term in the right-hand side of Eq.(\ref{eq15}) is reduced to the form
\begin{equation}
 \sqrt{-g} g^{\alpha\beta} \frac{\partial \Gamma^{\gamma}_{\alpha\gamma}}{\partial x^{\beta}} = \frac{\partial}{\partial x^{\beta}} \Bigl(\sqrt{-g} g^{\alpha\beta}
  \Gamma^{\gamma}_{\alpha\gamma}\Bigr) - \Gamma^{\gamma}_{\alpha\gamma} \frac{\partial (\sqrt{-g} g^{\alpha\beta})}{\partial x^{\beta}} \label{eq152}
\end{equation} 
From these three equations we obtain the explicit form of the vector $w^{\gamma}$ from Eq.(\ref{eq1})
\begin{equation}
  w^{\gamma} = g^{\alpha\beta} \Gamma^{\gamma}_{\alpha\beta} - g^{\beta\gamma} \Gamma^{\alpha}_{\alpha\beta} \label{eq153}
\end{equation}
and the following formula for the Lagrangian 
\begin{equation}
  {\cal L} = L \sqrt{-g} = \Gamma^{\gamma}_{\alpha\gamma} \frac{\partial (\sqrt{-g} g^{\alpha\beta})}{\partial x^{\beta}} - \Gamma^{\gamma}_{\alpha\beta} \frac{\partial 
 (\sqrt{-g} g^{\alpha\beta})}{\partial x^{\gamma}} - \sqrt{-g} g^{\alpha\beta} (\Gamma^{\rho}_{\alpha\gamma} \Gamma^{\gamma}_{\beta\rho} - \Gamma^{\gamma}_{\alpha\beta} 
  \Gamma^{\rho}_{\gamma\rho}) \label{eq154}
\end{equation}
where ${\cal L} = L \sqrt{-g}$ is the Lagrangian of the free gravitational field \cite{Carm2}. 

Now, by using a few formulas known for the derivatives which appear in Eq.(\ref{eq154}) (see, e.g., \cite{LLE}, \cite{Carm}) we can determine the difference of the first two terms 
from Eq.(\ref{eq154}) 
\begin{eqnarray}
  \Gamma^{\gamma}_{\alpha\gamma} \frac{\partial (\sqrt{-g} g^{\alpha\beta})}{\partial x^{\beta}} &-& \Gamma^{\gamma}_{\alpha\beta} \frac{\partial (\sqrt{-g} 
  g^{\alpha\beta})}{\partial x^{\gamma}} = 2 \sqrt{-g} g^{\rho\beta} \Gamma^{\gamma}_{\alpha\beta} \Gamma^{\alpha}_{\gamma\beta} - \sqrt{-g} g^{\beta\gamma} 
  \Gamma^{\alpha}_{\beta\gamma} \Gamma^{\rho}_{\alpha\rho} - \sqrt{-g} g^{\alpha\beta} \Gamma^{\gamma}_{\alpha\beta} \Gamma^{\rho}_{\gamma\rho} \nonumber \\
 &=& 2 \sqrt{-g} g^{\alpha\beta} (\Gamma^{\mu}_{\alpha\nu} \Gamma^{\nu}_{\beta\mu} - \Gamma^{\nu}_{\alpha\beta} \Gamma^{\mu}_{\nu\mu}) \label{eq155}
\end{eqnarray}
Therefore, the final expression for the Lagrangian ${\cal L}$ of the free gravitational field in metric GR is written in the form 
\begin{equation}
 {\cal L} = L \sqrt{-g} = \sqrt{-g} g^{\alpha\beta} ( \Gamma^{\mu}_{\alpha\nu} \Gamma^{\nu}_{\beta\mu} - \Gamma^{\nu}_{\alpha\beta} \Gamma^{\mu}_{\nu\mu})  \label{eq2}
\end{equation} 
This Lagrangian ${\cal L}$ is the Einstein-Hilbert Lagrangian of the metric GR. Note that in some old papers the quantity ${\cal L}$ was also called the Lagrangian density. The 
explicit form of ${\cal L}$, Eq.(\ref{eq2}), leads to the correct equations for all components of the gravitational field in metric GR. By using the explicit formulas for the 
Christoffel symbols $\Gamma^{\gamma}_{\alpha\beta}$ 
\begin{eqnarray}
   \Gamma^{\gamma}_{\alpha\beta} = \frac12 g^{\gamma\rho} \Bigl(\frac{\partial g_{\rho\alpha}}{\partial x^{\beta}} + 
   \frac{\partial g_{\rho\beta}}{\partial x^{\alpha}} - \frac{\partial g_{\alpha\beta}}{\partial x^{\rho}} \Bigr)  \label{eq3}
\end{eqnarray}
one reduces the formula, Eq.(\ref{eq2}), to the following `quadratic' form upon the derivatives of the metric tensor
\begin{eqnarray}
  {\cal L} &=& \frac14 \sqrt{-g} B^{\alpha\beta\gamma\mu\nu\rho} \Bigl(\frac{\partial g_{\alpha\beta}}{\partial x^{\gamma}}\Bigr) 
  \Bigl(\frac{\partial g_{\mu\nu}}{\partial x^{\rho}}\Bigr) \label{eq4} \\
  &=& \frac14 \sqrt{-g} (g^{\alpha\beta} g^{\gamma\rho} g^{\mu\nu} - g^{\alpha\mu} g^{\beta\nu} g^{\gamma\rho} + 2 g^{\alpha\rho} g^{\beta\nu} g^{\gamma\mu} - 
  2 g^{\alpha\beta} g^{\gamma\mu} g^{\nu\rho}) \frac{\partial g_{\alpha\beta}}{\partial x^{\gamma}} \frac{\partial g_{\mu\nu}}{\partial x^{\rho}} \nonumber
\end{eqnarray}
To simplify notations below the partial derivatives $\frac{\partial g_{\alpha\beta}}{\partial x^{\gamma}}$ are designated with the short notation $g_{\alpha\beta,\gamma}$. In this notation the 
Einstein-Hilbert Lagrangian, Eq.(\ref{eq2}), takes the form 
\begin{eqnarray}
  {\cal L} = \frac14 \sqrt{-g} B^{\alpha\beta\gamma\mu\nu\rho} g_{\alpha\beta,\gamma} g_{\mu\nu,\rho} \label{eq5}
\end{eqnarray}
where $B^{\alpha\beta\gamma\mu\nu\rho} = g^{\alpha\beta} g^{\gamma\rho} g^{\mu\nu} - g^{\alpha\mu} g^{\beta\nu} g^{\gamma\rho} + 2 g^{\alpha\rho} g^{\beta\nu} 
g^{\gamma\mu} - 2 g^{\alpha\beta} g^{\gamma\mu} g^{\nu\rho}$. 

This form of Einstein-Hilbert Lagrangian is used below to derive the actual Hamiltonian of metric GR. As follows from Eq.(\ref{eq5}) the Einstein-Hilbert Lagrangian is a cubic function of the 
components of the metric tensor $g^{\alpha\beta}$ and quadratic function of the derivatives of this tensor, i.e. it is quadratic upon the $\frac{\partial g_{\alpha\beta}}{\partial x^{\gamma}} 
= g_{\alpha\beta,\gamma}$ derivatives. It is important to note that each term in the Lagrangian ${\cal L}$, Eq.(\ref{eq5}), has the same structure. This fact simplifies the following analysis 
and formulation of the Hamilton approach. It should be mentioned that Lagrangian used in \cite{Dir58} (see also \cite{Fro2011}) is different from the Lagrangian ${\cal L}$, Eq.(\ref{eq5}). 
This is discussed in detail in Appendix A.

\section{Hamilton approach}

For actual applications of the Hamilton approach to the metric gravity we need to re-write the Einstein-Hilbert Lagrangian, Eq.(\ref{eq5}), to a slightly different form. Such a form must 
include explicit expressions for the temporal derivatives (or time-derivatives)  
\begin{eqnarray}
  {\cal L} = \frac14 \sqrt{-g} B^{\alpha\beta 0\mu\nu 0} g_{\alpha\beta,0} g_{\mu\nu,0} 
   + \frac12 \sqrt{-g} B^{(\alpha\beta 0|\mu\nu k)} g_{\alpha\beta,0} g_{\mu\nu,k} 
   + \frac14 \sqrt{-g} B^{\alpha\beta k \mu\nu l} g_{\alpha\beta,k} g_{\mu\nu,l} \label{eq51}
\end{eqnarray}
where the Latin indexes $k (\ge 1)$ and $l (\ge 1)$ designate the spatial coordinates only, while index 0 denotes the temporal coordinate, i.e. index 0 is the `temporal' index. In 
Eq.(\ref{eq51}) and everywhere below the notation $B^{(\alpha\beta\gamma|\mu\nu\rho)}$ means a tensor symmetrized in the two group of indexes, i.e.
\begin{eqnarray}
   B^{(\alpha\beta\gamma|\mu\nu\rho)} = \frac12 \Bigl( B^{\alpha\beta\gamma\mu\nu\rho} + B^{\mu\nu\rho\alpha\beta\gamma} \Bigr) \label{eq52}
\end{eqnarray}
Now note that in the classical Lagrange approach the Lagrangian ${\cal L}$ must be a function of all components of the metric $g_{\lambda\sigma}$ and their temporal derivatives 
$g_{\lambda\sigma,0}$ of the first order, i.e. ${\cal L} = {\cal L}(\{ g_{\lambda\sigma} \}, \{ g_{\lambda\sigma,0} \})$. The corresponding Lagrange equations are 
\begin{eqnarray}
  \frac{d}{d x_0}\Bigl(\frac{\partial {\cal L}}{\partial g_{\lambda\sigma,0}}\Bigr) = \frac{\partial {\cal L}}{\partial g_{\lambda\sigma}} \label{eq53}
\end{eqnarray}
In this study we do not want to discuss solution of these equations and their equivalence to the solutions of the Einstein's equations for metric GR (see e.g., \cite{Carm2} and Appendix B). Instead, 
let us consider the Hamilton approach to the metric GR which is also based on the Lagrangian, Eq.(\ref{eq51}). 

The first step in the Hamilton approach is the proper definition of all momenta. In our case these momenta are defined as the derivatives ${\cal L}$ upon the temporal derivatives of the components of 
the metric tensor, i.e. upon the $g_{\gamma\sigma,0}$ quantities. From Eq.(\ref{eq51}) one finds the explicit expressions for all momenta $\pi^{\gamma\sigma}$
\begin{eqnarray}
  \pi^{\gamma\sigma} = \frac{\partial {\cal L}}{\partial g_{\gamma\sigma,0}} = \frac12 \sqrt{-g} B^{((\gamma\sigma) 0|\mu\nu 0)} g_{\mu\nu,0} + \frac12 \sqrt{-g} 
  B^{((\gamma\sigma) 0|\mu\nu k)} g_{\mu\nu,k}  \label{eq54}
\end{eqnarray}
where the `double-symmetric' function $B^{((\alpha\beta)\gamma\mu\nu\rho)}$ is
\begin{eqnarray}
   B^{((\alpha\beta)\gamma|\mu\nu\rho)} = \frac14 \Bigl( B^{\alpha\beta\gamma\mu\nu\rho} + B^{\beta\alpha\gamma\mu\nu\rho} + B^{\mu\nu\rho\alpha\beta\gamma} 
  + B^{\mu\nu\rho\beta\alpha\gamma} \Bigr) \label{eq541}
\end{eqnarray}
It is straightforward to show that the function $B^{((\gamma\sigma) 0|\mu\nu 0)}$ equals to the product of the $g^{00}$ component and tensor $E^{\mu\nu\gamma\sigma}$ defined in \cite{Dir58}: 
$E^{\mu\nu\gamma\sigma} = e^{\mu\nu} e^{\gamma\sigma} - e^{\mu\gamma} e^{\nu\sigma}$, where $e^{\mu\nu} = g^{\mu\nu} - \frac{g^{0\mu} g^{0\nu}}{g^{00}}$. It is clear that, if either $\mu = 0$, or 
$\nu = 0$ (or both), then from these equalities one finds $e^{\mu\nu} = 0$ and $E^{\mu\nu\gamma\sigma} = 0$. Therefore, in such cases the $B^{((\gamma\sigma) 0|\mu\nu 0)}$ quantity in Eq.(\ref{eq54}) 
is singular. This means that from Eq.(\ref{eq54}) we cannot derive any analytical expression for the `velocities' $g_{0 \mu.0} = g_{\mu 0,0}$ and/or $g_{00,0}$ in terms of the corresponding momenta 
$\pi^{0\mu}$ and/or $\pi^{00}$. In other words, we are dealing with a constrained dynamical system \cite{Dir50}, \cite{Dir64}. In respect with the definition given in \cite{Dir50} and \cite{Dir64} 
all constraints which are directly related with the corresponding momenta are the primary constraints. The explicit form of these primary constraints in our case directly follows from Eq.(\ref{eq54}):
\begin{eqnarray}
  \phi^{0\sigma} = \pi^{0\sigma} - \frac12 \sqrt{-g} B^{((0\sigma) 0|\mu\nu k)} g_{\mu\nu,k}  \label{eq542}
\end{eqnarray}
It is easy to count that there are $d$ primary constraints in metric GR, where $d$ is the dimension of space-time.  

Consider now the regular case, i.e. when $\gamma\sigma = pq$ in Eq.(\ref{eq54}). In this case the `matrix' $B^{((pq) 0 \mu\nu 0)}$ in Eq.(\ref{eq54}) is invertible and for the 
`velocities' $g_{mn,0}$ one finds
\begin{eqnarray}
  g_{mn,0} = \frac{2}{\sqrt{-g} g^{00}} I_{mnpq} \pi^{pq} - \frac{1}{g^{00}} I_{mnpq} B^{((pq) 0|\mu\nu k)} g_{\mu\nu,k}  \label{eq543}
\end{eqnarray}
where the space-like or spatial) tensor $I_{mnpq}$ is inverse of $E^{pqkl}$, i.e. $I_{mnpq} E^{pqkl} = \delta^{k}_{m} \delta^{l}_{n} = E^{pqkl} I_{mnpq}$. The explicit formula for the $I_{mnpq}$
space-like tensor is \cite{Kuzm2008}
\begin{eqnarray}
  I_{mnpq} = \frac{1}{d - 2} g_{mn} g_{pq} - g_{mp} g_{nq} = \frac{1}{d - 2} g_{pq} g_{mn} - g_{pm} g_{nq} = I_{pqmn}  \label{eq544}
\end{eqnarray}
where the arising singularity at $d = 2$ corresponds to the one-dimensional gravity, when we have no `free' velocity and have to deal with a completely constrained motion \cite{Kuzm2008}.
    
By using the expression for the Lagrangian ${\cal L}$, Eq.(\ref{eq51}), and formulas for the `velocities' $g_{0\sigma,0}$ written in terms of momenta Eq.(\ref{eq542}) and Eq.(\ref{eq543}) we can 
obtain the following formula for the total Hamiltonian ${\cal H}_T$ of the metric gravity, or metric GR:
\begin{eqnarray}
  {\cal H}_T = \pi^{\alpha\beta} g_{\alpha\beta,0} - {\cal L} = {\cal H}_C + g_{0\sigma,0} \phi^{0\sigma}  \label{eq55}
\end{eqnarray}
where $\phi^{0\sigma}$ are the primary constraints, $g_{0\sigma,0}$ are the corresponding velocities and ${\cal H}_C$ is the canonical Hamiltonian of the metric GR
\begin{eqnarray}
 & &{\cal H}_C = \frac{1}{\sqrt{-g} g^{00}} I_{mnpq} \pi^{mn} \pi^{pq} - \frac{1}{2 g^{00}} I_{mnpq} B^{(m n 0|\mu \nu k)} g_{\mu\nu,k} \pi^{pq} \label{eq551} \\
 &-& \frac{1}{2 g^{00}} I_{mnpq} \pi^{mn} B^{(p q 0|\mu \nu k)} g_{\mu\nu,k} + \frac14 \sqrt{-g} \Bigl[ \frac{1}{g^{00}} I_{mnpq} B^{((mn)0|\mu\nu k)} B^{(pq0|\alpha\beta l)} 
 - B^{\mu\nu k \alpha\beta l}\Bigr] g_{\mu\nu,k} g_{\alpha\beta,l} \nonumber
\end{eqnarray}
These explicit forms of the total and canonical Hamiltonians ${\cal H}_T$ and ${\cal H}_C$ can be used in the future calculations and theoretical analysis of the metric GR. Note also that 
in the Hamilton approach we have two sets of the `conjugate' $d (d - 1)-$variables: $\frac{d (d - 1)}{2}$ `generalized' coordinates which are chosen as the components of the metric tensor 
$g_{\alpha\beta}$ and $\frac{d (d - 1)}{2}$ momenta $\pi^{\alpha\beta}$ conjugate to these coordinates. The classical Poisson brackets between these variables are  
\begin{eqnarray}
   [ g_{\alpha\beta}, \pi^{\mu\nu}] = - [ \pi^{\mu\nu}, g_{\alpha\beta}] = g_{\alpha\beta} \pi^{\mu\nu} - \pi^{\mu\nu} g_{\alpha\beta} = \frac12 \Bigl(\delta^{\mu}_{\alpha} 
   \delta^{\nu}_{\beta} + \delta^{\nu}_{\alpha} \delta^{\mu}_{\beta}\Bigr) = \Delta^{\mu\nu}_{\alpha\beta} \; \; \; ,  \label{eq552} 
\end{eqnarray}
where $\Delta^{\mu\nu}_{\alpha\beta}$ is the gravitational delta-function (or tensor delta-function). The Poisson bracket Eq.(\ref{eq552}) is the fundamental Poisson bracket, since  all other Poisson 
brackets equal zero identically, i.e. $[ g_{\alpha\beta}, g_{\mu\nu}] = 0$ and $[ \pi^{\alpha\beta}, \pi^{\mu\nu}] = 0$.  

Analytical derivation of the total and canonical (or dynamical) Hamiltonians $H_T$ and $H_C$ is the final step of the Hamilton procedure for any dynamical system. In the case of metric GR we are dealing 
with a constrained dynamical system to which we apply Dirac's procedure. Therefore, in this case we have to perform a few additional calculations. In particular, we need to show that the Dirac procedure 
is closed \cite{Dir64}. This means that the chain of constraints $[ \phi^{0\alpha}, H_C], [[ \phi^{0\alpha}, H_C], H_C], [[[ \phi^{0\alpha}, H_C], H_C], H_C], \ldots$ contains only a finite number of 
non-zero terms. To achieve this goal we note that the Poisson bracket (or PB, for short) of the primary constraints equals zero identically, i.e. $[ \phi^{0\alpha}, \phi^{0\beta}] = 0$. Second crucial 
fact follows directly from the canonical Hamiltonian $H_C$, Eq.(\ref{eq551}), which does not include any momentum $\pi^{0\alpha}$. This means that the Poisson bracket $[ \phi^{0\alpha}, H_C]$ cannot 
contain any primary constraint. In other words, if the Poisson bracket $[ \phi^{0\alpha}, H_C]$ is not equal zero, then it is proportional to the secondary constraint(s) $\chi^{0\alpha}$. The explicit 
formulas for all secondary constraints $\chi^{0\alpha}$ have been found in \cite{Kuzm2008}. These expressions for $\chi^{0\alpha}$ are cumbersome and here we do not want to repeat them. Note only that 
the Poisson bracket between secondary constraints $\chi^{0\alpha}$ and canonical Hamiltonian $H_C$ are represented as a linear combination of the same secondary constraints $\chi^{0\alpha}$ (see, Eq.(17) 
in \cite{Kuzm2008}). The coefficients of such linear combination are the field-dependent functions. Formally, this shows the closure of the Dirac procedure, since no tertiary constraints arises in the 
metric GR. The Poisson brackets between the primary and secondary constrains are: $[ \phi^{0\alpha}, \chi^{0\beta}] = \frac12 g^{\alpha\beta} \chi^{00}$, i.e. all of them proportional to the secondary 
constraint $\chi^{00}$. 
 
\section{Application of the Hamilton approach}

The Hamilton approach can be applied to a number of actual problems currently known in metric GR. First, following \cite{Kuzm2008} one can show that both Lagrange and Hamilton approaches lead to the 
equations of motion which are invariant under the diffeomorphism transformation. For the Lagrange approach this was known for quite some time, while for the Hamilton approach this has been shown only 
recently \cite{Kuzm2008}. By using the well known Castellani procedure \cite{Castel} to derive generators of the gauge transformations one can show \cite{Kuzm2008} that the Hamilton approach gives the 
same result for the gauge invariance of the metric GR (diffeomorphism) as it follows from the Lagrange approach \cite{Saman}. This result is one of the great achievements of the Dirac procedure 
\cite{Kuzm2008}, since all previously developed, `alternative' Hamiltonian-based formulations of the metric GR, including the notorious ADM Hamiltonian formulation, could not reproduce this simple and 
obvious result (see discussions in \cite{Kuzm2008}, \cite{Fro2011} and \cite{Kuzm2009}). 

Second, there is an explicit expression for the canonical Hamiltonian $H_C$, Eq.(\ref{eq551}), written in terms of the secondary constraints and an additional surface term. For the Hamilton approach 
considered here this formula takes the form
\begin{eqnarray}
  H_C = - 2 g_{0\lambda} \chi^{0\lambda} + \Bigl[ 2 g_{0m} \pi^{mk} - \sqrt{-g} \Bigl( g_{0n} B^{((nk) 0|\alpha\beta l)} + g_{0\gamma} B^{((0\gamma) k|\alpha\beta l)} 
  g_{\alpha\beta,l} \Bigr) \Bigr]_{,k} \label{HamC} 
\end{eqnarray}
Analogous expression for the $H_C$ Hamiltonian can be derived (see, e.g., \cite{Fro2011}) by applying the Dirac's formulation of the metric GR \cite{Dir58}. Note that often the formula, Eq.(\ref{HamC}), 
is used to illustrate some  weakness of the Hamilton approach which arise after application of the Dirac procedure for constrained dynamical systems. Indeed, it is hard to assume $a$ $priori$ that the 
total energy of the free gravitational field in a closed spatial volume which contains no gravitational sources (i.e. masses) is always a constant. However, as follows from the formula, Eq.(\ref{HamC}), 
such an energy is always a constant unless we have a non-zero gravitational flux through the borders of this closed volume $V$. Let us consider the surface term in the formula, Eq.(\ref{HamC}), more 
carefully.   

As follows from Eq.(\ref{HamC}) the canonical and total Hamiltonians $H_{C}$ and $H_T$ are the sum of the terms proportional to the secondary $\chi^{0\lambda}$ constraints and a surface term which is a 
combination of the total spatial derivatives. This surface term can be represented in a slightly different form with the use of the following spatial vector (or $(d-1)-$vector) $\overline{G} = 
(G^1, G^2, \ldots, G^d)$, where 
\begin{eqnarray}
 G^{k} =  g_{0m} \pi^{mk} - \sqrt{-g} \Bigl[ g_{0n} B^{((nk) 0|\alpha\beta l)} + g_{0\gamma} B^{((0\gamma) k|\alpha\beta l)} g_{\alpha\beta,l} \Bigr]_{,k} \label{vector}
\end{eqnarray}
is the $k$th contravariant component of this $(d - 1)-$vector ($k = 1, 2, \ldots, d - 1$). The $(d-1)-$vector $\overline{G}$ is the energy flux of the free gravitational field, i.e. this vector determines 
the flow of the gravitational energy (or, gravitational flow, for short) through the closed boundary $(d-1)-$surface of the volume occupied by the free gravitational field only. No sources of gravitation,
e.g., masses, can be located in this volume. By calculating the integral from the left-hand side of Eq.(\ref{eq55}) over the whole volume $V = V_d$ occupied by the free gravitational field and enclosed 
by the closed surface $S_{d-1}$ one finds   
\begin{eqnarray}
 \Delta E = \int div \overline{G} \cdot dV_d = - \oint (\overline{G} \cdot \overline{n}) dS_{d-1} = - \oint \overline{G} \cdot d\overline{S}_{d-1}  \label{gauss}
\end{eqnarray}
where $E$ is the total energy of the gravitational field in the finite volume $V$ ($E = \pi^{\alpha\beta} g_{\alpha\beta} - {\cal L}$, see Eq.(\ref{eq55}) above), $\overline{G}$ is the (d-1)-dimensional 
vector defined in Eq.(\ref{vector}), $\overline{n}$ is the unit vector of the outer normal to the surface element $dS_{d-1}$ and $d\overline{S}_{d-1} = \overline{n} dS_{d-1}$ is the elementary volume 
of the surface $dS_{d-1}$ oriented in the direction of the outwardly directed normal $\overline{n} = (n_1, n_2, \ldots, n_d)$. To transform the integral in Eq.(\ref{gauss}) we have applied the Gauss 
formula for multi-dimensional integrals. Now it is clear that the formulas Eq.(\ref{HamC}), Eq.(\ref{vector}) and Eq.(\ref{gauss}) represent the fact that the total energy of the free gravitational field 
in a finite volume $V$ (which is free from any actual mass) is a constant unless some non-zero flux of gravitational energy crosses the surface $S_{d-1}$ of this volume $V$. In this case the change of 
the total energy $E$ is governed by Eq.(\ref{gauss}). This fact is directly related to the properties of the actual gravitational filed in metric GR, rather than with some defect of the Dirac's procedure. 
To avoid additional questions about actual propagation of the free gravitation field it is better to restrict applications of formulas Eq.(\ref{vector}) and Eq.(\ref{gauss}) to the case of very small, or 
even infinitely small (infinitesimal), volume $V$.   

The spatial vector $\overline{G}$ in Eq.(\ref{vector}) plays the same role in metric General Relativity as the Pointing vector plays in Electrodynamics \cite{LLE}. Note that the left-hand side of the 
energy conservation law must contain the time-derivative of the total field energy, i.e. $\frac{\partial E}{\partial t}$. The same general identity must be correct in the metric GR. Now, let us assume that
the energy of propagating gravitational wave is located in its front. In this case the expression for $\delta E$ (see, Eq.(\ref{gauss})) can be transformed in the following way 
\begin{eqnarray}
 \Delta E &=& \int_{t}^{t + \Delta t} \frac{\partial E}{\partial t} dt = \int_{t}^{t + \Delta t} \frac{\partial w}{\partial t} dt dV = \int_{t}^{t + \Delta t} \oint \frac{\partial w}{\partial t} 
  \frac{v}{c} c dt dS_{d-1} \nonumber \\
 &\approx& \frac{v_f}{c} \int \oint \Bigl(\frac{\partial w}{\partial t}\Bigr) c dt dS_{d-1} = E_f \label{gauss1} 
\end{eqnarray}
where $E_f$ is the energy located at the front of the propagating gravitational wave, $w$ is the spatial density of the energy $E$, i.e. the energy per unit volume, i.e. $w = \lim_{V \rightarrow 0} \Bigl( 
\frac{E}{V} \Bigr)$, while $c$ is the speed of light in vacuum and $v_f$ is the propagation velocity of the gravitational wave in vacuum. Note that from Eq.(\ref{gauss1}) we have $\Delta E = E_f$, i.e. the 
energy of the propagating gravitational wave is located in the frontal area of such a wave only. Also, the time $\Delta t$ in this formula coincides with the time when the propagating gravitational wave 
crosses the boundary surface $S_{d-1}$. Very likely, that the velocity of the front propagation $v_f$ equals to the speed of light in vacuum exactly, i.e. $v_f = c$. However, such an assumption must be 
confirmed in a number of independent experiments. Everywhere below in this study we shall assume that $v_f = c$. 

Thus, to determine the gravitational energy of the free gravitational field(s) one needs to answer the fundamental question about propagation of gravitation in the metric GR. Formally, we can say that 
gravitational fields propagate by the `gravitational waves', but physical meaning of these `waves' becomes clear only if we know their structure and propagation laws. In reality, the internal structure 
of the gravitational waves can be investigated by using the explicit form of the total and canonical Hamiltonians, Eq.(\ref{eq55}) and Eq.(\ref{eq551}). Let us restrict ourselves to the analysis of the
canonical Hamiltonian ${\cal H}_C$, Eq.(\ref{eq551}). Furthermore, in this study we can try to answer only one question about pure harmonic oscillations for the gravitational fields, or in other words, 
about gravitational waves propagating in space as light waves, or waves generated by a set of harmonic oscillators. In early years of the metric GR Einstein shown \cite{E1918} that very weak 
gravitational field(s) can propagate in space as harmonic vibrations of the constant frequencies. However, later he and his co-workers considered actual gravitational fields, which are not very 
weak, \cite{E1937} and arrived to a very different conclusion which can be formulated as follows: \textit{propagation of the gravitational waves in the form of pure harmonic vibrations is not possible 
unless the gravitational filed is very weak}. Such a conclusion follows from the explicit form of the Hamiltonian, Eq.(\ref{eq551}). To show this in detail let us formulate the principal question about 
gravitational waves in a slightly different form by assuming that we have found a canonical Hamiltonian $H_H$ for the gravitational field which describes pure harmonic oscillations of this field. In 
addition to this the leading part of such a Hamiltonian ($\simeq \pi^{mn} \pi^{pq}$) coincide with the corresponding part of the canonical Hamiltonian $H_C$, Eq.(\ref{eq551}):    
\begin{eqnarray}
   {\cal H}_{H} = \frac{1}{\sqrt{-g} g^{00}} I_{mnpq} \Bigl( \pi^{mn} \pi^{pq} + \Omega^{klrt}_{mnpq} g_{kl} g_{rt} \Bigr) \; \; \; , \; \; \;  \label{HarHam}
\end{eqnarray}  
where $\Omega^{klrt}_{mnpq}$ is a spatial $4 \times 4$ tensor. To describe pure harmonic oscillations this tensor must be: (1) truly $g-$independent, i.e. independent of all components of the metric 
tensor $g_{\alpha\beta}$, and (2) positively defined, i.e. all its eigenvalues must be positive. If some of the components of the spatial $\Omega^{klrt}_{mnpq}$ tensor are $g-$dependent, then the 
profile of the gravitational wave will be changed during its propagation. If any of these conponents is negative, then the amplitude of the gravitational wave will increase during its propagation. 

Now consider the actual Hamiltonian for metric GR, i.e. the quantity $H_C$ defined by Eq.(\ref{eq551}). The fundamental question is: can we reduce the actual Hamiltonian $H_C$ to the form of harmonic 
Hamiltonian written in Eq.(\ref{HarHam}) with some positively defined spatial tensor $\Omega^{klrt}_{mnpq}$ which does not depend upon components of the metric tensor $g_{\alpha\beta}$? It can be shown 
that the answer to this question is negative, since the tensor $\Omega^{klrt}_{mnpq}$ in Eq.(\ref{HarHam}) is substantially $g-$dependent. The word `substantially' is used here to emphasize that by using 
only linear transformations of the metric components and corresponding momenta we cannot reduce this tensor to a `constant' spatial tensor. As a maximum we can transform the Hamiltonian, Eq.(\ref{eq551}), 
to the following form
\begin{eqnarray}
   {\cal H}_{C} = \frac{1}{\sqrt{-g} g^{00}} I_{mnpq} \Bigl\{ \pi^{mn} \pi^{pq} + \Bigl[ P^{klrt}_{mnpq}(\{ g_{\alpha\beta} \}) + \sqrt{-g} G^{klrt}_{mnpq}(\{ g_{\alpha\beta} \}) \Bigr] 
   g_{kl} g_{rt} \Bigr\} \; \; \; , \; \; \;  \label{HamCH}
\end{eqnarray}  
where $P^{klrt}_{mnpq}(\{ g_{\alpha\beta} \})$ and $G^{klrt}_{mnpq}(\{ g_{\alpha\beta} \})$ are the two polynomial-type functions of all components of the metric tensor. The maximal power of these 
polynomials equals four. Another source of non-linearity is the $\sqrt{-g}$ factor which cannot be replaced by some numerical constant (e.g., unity) for non-weak gravitational filed(s). Since we cannot 
reduce the Hamiltonian Eq.(\ref{HamCH}) to the form of Eq.(\ref{HarHam}), then we have to conclude that there are no gravitational wave which propagate as pure harmonic vibrations, or harmonic waves, for 
short. It is also clear that there is no need to discuss the positive definition of the tensor $\Omega^{klrt}_{mnpq}$ in the metric gravity, since this tensor and all its eigenvalues are substantially 
$g-$dependent. This result is of fundamental importance for the metric General Relativity. In particular, it indicates clearly that the free gravitational fields cannot propagate in a space-time 
continuum as `harmonic vibrations' (or oscillations). 

In the case of very weak gravitational fields one can find a similarity with the free electromagnetic field. Indeed, for very weak gravitational fields the differences between the corresponding components 
of the metric tensor and Minkovskii tensor are small and $\sqrt{-g} = 1$. In this case the Hamiltonians Eq.(\ref{eq551}) and Eq.(\ref{HamCH}) are the quadratic functions of the new variables 
$h_{\alpha\beta}$ an momenta conjugate to them $\pi_{\gamma\rho}$, where $h_{\alpha\beta} = g_{\alpha\beta} - \eta_{\alpha\beta}$ are the small corrections to the corresponding components of the Minkowskii 
tensor $\eta_{\alpha\beta} = diag(-, +, +, \ldots, +)$ in the flat space-time. Formally, this means that very weak gravitational fields can propagate as `harmonic' vibrations with the `constant' frequencies
and amplitudes. In this case we have an obvious similarity with the propagation of the free electromagnetic fields. 

For arbitrary gravitational fields we have $\sqrt{-g} \ne 1$ and the values $h_{\alpha\beta} = g_{\alpha\beta} - \eta_{\alpha\beta}$ are not small. In this case we are back to the gravitational Hamiltonians 
represented by Eq.(\ref{eq551}) and Eq.(\ref{HamCH}). The $g-$dependence of the $\Omega^{klrt}_{mnpq}$ spatial tensor in Eq.(\ref{HarHam}) leads to substantial changes in the profile of the propagating 
gravitational waves. In particular, the amplitude of such a propagating wave rapidly increases at its front when wave propagates. Finally, the front of the propagating gravitational wave will contain all the
energy of this wave. This conclusion about internal structure of the propagating gravitational wave follows from our analysis of the Hamiltonian, Eq.(\ref{eq551}), of the metric GR.
     
It follows from here that all energy of the propagating gravitational wave is associated only with the front of such a wave. Before and after the wave front area the local gravitational energy, i.e. energy 
spatial density, is a constant which can be equal zero. This conclusion follows from the fact that the total Hamiltonian is zero before the front of the propagating gravitational wave and it equals to the 
sum of constraints (i.e. zero) after the wave front. The only non-zero term in the total Hamiltonian $H_T$ is the surface term which describes the gravitational flow through the surface which has been 
reached by a propagating gravitational wave. The concentration of the whole energy of the propagating gravitational wave in its front is the direct consequence of substantial non-linearity of the field 
equations in metric GR. In some sense the propagating gravitational wave is similar to a very strong shock wave which propagates in a compressible gas mixture. However, in contrast with the shock wave(s) in
gas dynamics we cannot apply any discontinuity condition to the gravitational wave(s). Therefore, we cannot discuss any `rarefaction' (or back) front of the propagating gravitational wave. This is a brief 
description of the internal structure of the propagating gravitational wave. Such a structure is relatively simple, e.g., it does not include any oscillations, but it is clear that only this structure agrees 
with the original ideas of GR proposed and developed by Mach and Einstein. In actual applications we always have sequences of the propagating gravitational waves. 

The last question which we want to discuss here is related to the quantization of the gravitational field in metric GR. Below we follow to the procedure developed by Dirac for constrained dynamical systems 
\cite{Dir50} (see also \cite{Dir64}, \cite{Tyut}, \cite{Fro2013}). To perform quantization of the metric GR we need to replace the classical Poisson bracket $[ \pi^{\alpha\beta}, g_{\mu\nu}]$ by the 
corresponding quantum Poisson bracket. The classical Poisson bracket between two quantities $A$ and $B$ is defined traditionally
\begin{eqnarray}
   [ A, B ]_{Cl} = [ A, B ]^{\mu\nu}_{\alpha\beta} = \frac{\partial A}{\partial g^{\alpha\beta}} \frac{\partial B}{\partial \pi^{\mu\nu}} - \frac{\partial A}{\partial \pi^{\mu\nu}} 
  \frac{\partial B}{\partial g_{\alpha\beta}} \label{PBCl}
\end{eqnarray}
where the index $Cl$ is used for the word `Classic'. To write the explicit expression for the quantum Poisson bracket we have to remember that in Quantum Mechanics the momenta $\pi^{\alpha\beta}$ and 
coordinates $g_{\mu\nu}$ cannot be measured simultaneously in one space-time point. This means that we have to chose a different representation of these variables in Quantum Mechanics. The simplest way 
is to assume that we can measure all generalized `coordinates' $g_{\mu\nu}$. This leads to the following representation of the momenta $\pi^{\alpha\beta}$ by the differential operators, i.e. 
\begin{equation}
 \pi^{\mu\nu} = - \imath \hbar \Bigl[ \frac{\partial}{\partial g_{\mu\nu}} + f_{\mu\nu}(g_{\alpha\beta}) \Bigr] \label{Qmom}
\end{equation}
where $f_{\mu\nu}(g_{\alpha\beta})$ is a regular (or analytical) function of all components of the metric tensor. To derive Eq.(\ref{Qmom}) we have used the fact that the quantum Poisson bracket must 
explicitly contain the reduced Planck constant $\hbar$ which is an universal measure of `non-commutativity' of different operators in Quantum Mechanics. The rest of the formula, Eq.(\ref{Qmom}), can be 
found by applying the `correspondence principle' known in Quantum Mechanics since the middle of 1920's (see, e.g., \cite{DirQuant} and \cite{LLQ}). For the free gravitational filed the correspondence 
principle means that the 
quantum Poisson bracket must have the correct limit in the case of very weak gravitational fields and for the Cartesian coordinates, or, in other words, for the flat space-time. This determines the 
following expression for the quantum (Q) Poisson bracket between coordinates and momenta
\begin{eqnarray}
 [ g_{\alpha\beta}, \pi^{\mu\nu}]_Q = \imath \hbar \frac12 ( \delta^{\mu}_{\alpha} \delta^{\nu}_{\beta} + \delta^{\nu}_{\alpha} \delta^{\mu}_{\beta} ) = \imath \hbar \Delta^{\mu\nu}_{\alpha\beta} 
 \label{PoisQ}
\end{eqnarray}   
This formula agrees with Eq.(\ref{Qmom}) for the quantum operator of momentum $\pi^{\alpha\beta}$ in the $g_{\alpha\beta}$-representation, or in the `coordinate' representation.

The functions $f_{\mu\nu}(g_{\alpha\beta})$ in Eq.(\ref{Qmom}) are the regular (or analytical) functions which depend upon all components of the metric tensor. The quantum PB between two arbitrary 
coordinates and two arbitrary momenta must be equal zero identically. From here one finds a number of additional conditions for the $f_{\mu\nu}$-functions from Eq.(\ref{Qmom}) 
\begin{equation}
 \frac{\partial f_{\mu\nu} }{\partial g_{\alpha\beta} } = \frac{\partial f_{\alpha\beta} }{\partial g_{\mu\nu}} \label{cond1}
\end{equation}
In general, one can use some freedom to choose different types of the $f_{\mu\nu}$ functions in Eq.(\ref{Qmom}) to simplify either the definition of momenta $\pi^{\mu\nu}$, or the formula for the quantum 
Hamiltonian operators $H^{Q}_{T}$ and $H^{Q}_{C}$ which are derived from the classical Hamiltonian operators $H_{T}$, Eq.(\ref{eq55}), and $H_{C}$, Eq.(\ref{HamC}). For instance, if we chose all 
$f_{\mu\nu}$ functions in Eq.(\ref{Qmom}) equal zero identically, then the quantum Hamiltonian $H^{Q}_{C}$ takes the form 
\begin{eqnarray}
 &{\cal H}^{Q}_C& = -\hbar^2 \frac{1}{\sqrt{-g} g^{00}} I_{mnpq} \frac{\partial^2}{\partial g_{pq} \partial 
 g_{mn}} - \frac{\hbar}{g^{00}} I_{mnpq} B^{(m n 0|\mu \nu k)} g_{\mu\nu,k} \frac{\partial}{\partial g_{pq}} \nonumber  \\
 &+& \frac{\hbar}{2 g^{00}} I_{mnpq} \Bigl[ \frac{\partial}{\partial g_{mn}}, B^{(p q 0|\mu \nu k)} g_{\mu\nu,k} \Bigr] \label{eq551Q} \\
 &+& \frac14 \sqrt{-g} \Bigl[ \frac{1}{g^{00}} I_{mnpq} B^{((mn)0|\mu\nu k)} B^{(pq0|\alpha\beta l)} - B^{\mu\nu k \alpha\beta l}\Bigr] g_{\mu\nu,k} g_{\alpha\beta,l} \nonumber
\end{eqnarray}
The corresponding Schr\"{o}dinger equation for the free gravitational field is written in the form 
\begin{eqnarray}
   \imath \hbar \frac{\partial \Psi}{\partial t} = {\cal H}^{Q}_{T} \Psi = {\cal H}^{Q}_{C} \Psi \label{eq555Q}
\end{eqnarray}
with a set of $2d-$additional conditions: $\phi^{0\sigma} \Psi = 0$ and $\chi^{0\sigma} \Psi = 0$ for the wave function $\Psi$ which depends upon all components of the metric tensor, i.e. $\Psi = 
\Psi(g_{00}, g_{01}, \ldots, g_{dd})$. Here $\phi^{0\sigma}$ and $\chi^{0\sigma}$ are the primary and secondary constraints mentioned above written in their operator forms (or quantum forms). If the 
Schr\"{o}dinger equation for the free gravitational field is written with the use of the variable $x_0 = c t$ instead of $t$, where $c$ is the speed of light in vacuum, then an additional factor $c$ 
must be used in the left-hand side of this equation. As expected (see discussion in \cite{Heitl}) neither the Schr\"{o}dinger equation for the free gravitational field, nor Poisson brackets 
Eq.(\ref{PoisQ}) include any constant associated with the structure of matter. This means that the gravitational constant $k$, or any particle mass can be included in these equations which contain 
only $\hbar$ and $c$ (or $\hbar, c$ and $v_f$) as for the free electromagnetic field \cite{Heitl}. In conclusion, it should be mentioned that theoretical derivation of the Schr\"{o}dinger equation and 
additional conditions for the free gravitational field is not a difficult problem, but any observation of the effects directly related with quantum (metric) gravity is still far away from our current 
experimental abilities. To illustrate this we only note that the Compton wavelength of our Sun is very small ($< 1 \cdot 10^{-60}$ $cm$) and currently nobody knows when and how we can study such short 
distances in experiments. 

\section{Conclusions}

We have considered the Hamilton approach to the metric General Relativity (or metric GR, for short) of the free gravitational field. It is shown that the Hamilton approach can be developed in a short 
and transparent way which starts from the original Lagrangian, Eq.(\ref{eq154}), which was used in the Hilbert gravitational action $S_g$, Eq.(\ref{eq1}). Note that our results coincide with the results 
obtained in earlier studies where the same problem was considered \cite{Pir52} and \cite{Kuzm2008}. This is a strong indication that finally the correct Hamilton approach has been developed for the 
metric gravity. This approach can now be used to solve a number of actual problems which currently exist in the metric GR.    

The main result of this study is the analytical, Hamiltonian-based description of the free gravitational field which is free from internal contradictions. Dynamical variables in this approach are the 
components of the metric tensor $g_{\alpha\beta}$ (`coordinates') and momenta $\pi^{\mu\nu}$ conjugate to them. With such a choice of dynamical variables the free gravitational field becomes a `natural' 
dynamical system. Indeed, the Lagrangian of this system is a homogeneous quadratic functions upon velocities, while both total and canonical Hamiltonians are also quadratic functions of the space-like 
components of momentum, i.e. the $\pi^{mn}$ variables. The canonical Hamiltonian is written in a closed form in terms of pairs of the conjugate variables $\pi^{mn}$ and $g_{kl}$. In addition to this there 
are also $d-$primary and $d-$secondary constraints (or gauge conditions) for the free gravitational field. Based on the derived Hamiltonian we consider the problem of quantization of the free gravitational 
field and derive the corresponding Schr\"{o}dinger equation for this field. When quantization is finished the $2d-$constraints are written as additional gauge conditions for the wave function of the field. 
We also discuss the internal structure of the propagating gravitational wave. 

It should be mentioned in conclusion that there is a similarity between Lagrangians of the free electro-magnetic and free gravitational fileds. This leads to similarity between the corresponding total 
Hamiltonians constructed for these fields. Indeed, the free electro-magnetic and free gravitational fileds have only primary and secondary constraints (no tertiary constraints) and the total number of
primary and secondary constraints equals for both fields (one primary and one secondary constraints for the free electro-magnetic field and $d$-primary and $d-$secondary constraints for the free 
gravitational field). In other words, each primary constraints generates only one secondary constraint. Canonical Hamiltonians for both fields are quadratic functions upon space-like components of momemta. 
There are a few other similarities but we do not want to discuss them here. 

\section{Acknowledgments}

I am grateful to my friends D.G.C. (Gerry) McKeon, N. Kiriushcheva and S.V. Kuzmin (all from the University of Western Ontario, London, Ontario, CANADA) for helpful discussions and inspiration.

\begin{center}
  {\bf Appendix A}
\end{center}

The difference between Hamilton approach discussed in this study (see also \cite{Pir52}, \cite{Kuzm2008}) and Hamilton approach developed in \cite{Dir58} (see also \cite{Fro2011}) follows from the fact 
that Dirac used a different Lagrangian for the metric GR. The difference between these Lagrangians can be understood from the following simple relation \cite{Fro2011}
\begin{eqnarray}
  {\cal L}_{Dir} &=& {\cal L}_{PSS} - \Phi = {\cal L}_{PSS} - \Bigl[ (\sqrt{-g} g^{00})_{,k} \frac{g^{0k}}{g^{00}}\Bigr]_{,0} + \Bigl[ (\sqrt{-g} g^{00})_{,0} \frac{g^{0k}}{g^{00}}\Bigr]_{,k} \nonumber \\
  &=& {\cal L}_{PSS} - \Bigl[ (\sqrt{-g} g^{00})_{,\alpha} \frac{g^{0\alpha}}{g^{00}}\Bigr]_{,0} + \Bigl[ (\sqrt{-g} g^{00})_{,0} \frac{g^{0\alpha}}{g^{00}}\Bigr]_{,\alpha} \label{DirPir}  
\end{eqnarray}
where ${\cal L}_{PSS}$ is the Lagrangian of the metric GR used by Pirani et al \cite{Pir52} and above (see also \cite{Kuzm2008}), while ${\cal L}_{Dir}$ is the Lagrangian of the metric GR used by Dirac 
(see also \cite{Fro2011}). As follows from Eq.(\ref{DirPir}) the explicit expression for the function $\Phi(\{ g_{\mu\nu} \})$ is
\begin{eqnarray}
  \Phi &=&  \Bigl[ (\sqrt{-g} g^{00})_{,\alpha} \frac{g^{0\alpha}}{g^{00}}\Bigr]_{,0} - \Bigl[ (\sqrt{-g} g^{00})_{,0} \frac{g^{0\alpha}}{g^{00}}\Bigr]_{,\alpha}  \nonumber \\
  &=& \Bigl(\sqrt{-g} g^{00}\Bigr)_{,\alpha} \Bigl[\frac{g^{0\alpha}}{g^{00}}\Bigr]_{,0} - \Bigl(\sqrt{-g} g^{00}\Bigr)_{,0} \Bigl[\frac{g^{0\alpha}}{g^{00}}\Bigr]_{,\alpha} \label{DirPir1}  
\end{eqnarray}
It is easy to show that the integral over any finite area of space-time continuum from the last expression in Eq.(\ref{DirPir1}) equals zero identically, i.e. we do not change any `observable' property 
of the system/field. The function $\Phi$ is the function which generates the canonical transformation of variables. This transformation leads to a different definition of momenta and  different expressions 
for the primary constraints \cite{Fro2011}. All secondary constraints and other relations change correspondingly \cite{Fro2011}. The function $\Phi = \Phi(\{ g_{\mu\nu} \})$ allows one to make a canonical 
transition from the $H_T$ and $H_C$ Hamiltonians derived in \cite{Pir52}, \cite{Kuzm2008} to the $H_T$ and $H_C$ Hamiltonians which are obtained in \cite{Dir58} and \cite{Fro2011}. It is clear that it is 
possible to construct a large number (in principle, infinite number) of other `generating functions', i.e. functions which generate different sets of canonical transformations in metric GR.

\begin{center}
  {\bf Appendix B}
\end{center}

The formula Eq.(\ref{eq51}) leads to the following formula for the second variation of the Lagrangian ${\cal L}$
\begin{eqnarray}
   \frac{\partial^2 {\cal L}}{\partial g_{\alpha\beta,0} \partial g_{\mu\nu,0}} = \frac14 \sqrt{-g} B^{((\alpha\beta)0|(\mu\nu)0)} = \frac18 \sqrt{-g} \Bigl[ B^{((\alpha\beta) 0|\mu\nu 0)}
   + B^{((\alpha\beta) 0|\nu\mu 0)} \Bigr] \label{Leg1}
\end{eqnarray}
By assuming the Lagrangian ${\cal L}$ of the metric gravity has a minimum along the actual extremal and by applying the Legendre condition (see, e.g., Chpts. 5 and 6 in \cite{Gelf}) for this Lagrangian, 
one finds that the following inequality 
\begin{eqnarray}
   \frac18 \sqrt{-g} B^{((\alpha\beta)0|(\mu\nu)0)} \ge 0 \; \; , \;  or \; \; \; B^{((\alpha\beta) 0|\mu\nu 0)} + B^{((\alpha\beta) 0|\nu\mu 0)} \ge 0 \label{Leg2}
\end{eqnarray}
must always be obeyed for actual gravitational extremals. Here and below the notation $A^{\alpha\beta\gamma\rho} \ge 0$ means that the tensor $A^{\alpha\beta\gamma\rho}$ is positively defined, i.e. 
all its eigenvalues are non-negative (they are positive or equal zero). The last inequality can also be written in the form
\begin{eqnarray}
   B^{((\alpha\beta) 0|\mu\nu 0)} + B^{((\alpha\beta) 0|\nu\mu 0)} = g^{00} \Bigl( E^{\mu\nu\alpha\beta} + E^{\mu\nu\beta\alpha} \Bigr) \ge 0 \label{Leg25}
\end{eqnarray}
where the tensor $E^{\mu\nu\alpha\beta} = e^{\mu\nu} e^{\alpha\beta} - e^{\mu\alpha} e^{\mu\beta}$ has been defined by Dirac \cite{Dir58} and $e^{\mu\nu} = g^{\mu\nu} - \frac{g^{0\mu} g^{0\nu}}{g^{00}}$.
This leads to the following inequality
\begin{eqnarray}
   g^{00} \Bigl[ e^{\alpha\beta} e^{\mu\nu} - \frac12 \Bigl( e^{\alpha\mu} e^{\beta\nu} + e^{\alpha\nu} e^{\beta\mu} \Bigr) \Bigr] \ge 0 \label{Leg3}
\end{eqnarray}
This inequality can be useful in metric GR. In particular, from Eq.(\ref{Leg2}) one finds
\begin{eqnarray}
   g^{00} \Bigl[ e^{km} e^{pq} - \frac12 \Bigl( e^{kp} e^{mq} + e^{kq} e^{mp} \Bigr) \Bigr] > 0 \label{Leg5}
\end{eqnarray}
which means that the product of $g^{00}$ and following space-like tensor is positively defined, i.e. all its eigenvalues are positive. This product plays a role of the always positive `mass' (or
`mass-tensor') for the free gravitational field in metric GR. Another group of useful inequalities is discussed in the Appendix C.  

\begin{center}
  {\bf Appendix C}
\end{center}

In addition to the equalities mentioned above there is a second group of inequalities for the components of metric tensor $g^{\alpha\beta}$ and the corresponding momenta $\pi_{\alpha\beta}$. These 
inequalities directly follow from the Young inequality for the sum of the Lagrangian and Hamiltonian of the free gravitational field in metric GR. Investigation of this problem leads to the 
formulation of basic variational principles which can be used in calculations of the actual gravitational fileds. Derivation of this fundamental inequality for constrained systems requires an additional 
explanation.  First, consider an idealized situation when gravitational fields in the metric GR can be considered as regular dynamical systems, i.e. dynamical systems which have no constraints. In this 
case transition from the Lagrangian ${\cal L}$ to the total Hamiltonian $H_T$ can be performed with the use of the (convex) Legendre transformation 
\begin{eqnarray}
 H_T(x_0, g_{\alpha\beta}, \pi^{\alpha\beta}) = \max_{g_{\alpha\beta,0}} \Bigl[ g_{\alpha\beta,0} \pi^{\alpha\beta} - {\cal L}(x_0, g_{\alpha\beta}, g_{\alpha\beta,0}) \Bigr] \label{Legen1}
\end{eqnarray}  
where all notations are the same as in the main text. Since the convex Legendre transformation is an involution, then we can also write 
\begin{eqnarray}
  {\cal L}(x_0, g_{\alpha\beta}, g_{\alpha\beta,0}) = \max_{\pi^{\alpha\beta}} \Bigl[ g_{\alpha\beta,0} \pi^{\alpha\beta} - H_T(x_0, g_{\alpha\beta}, \pi^{\alpha\beta}) \Bigr] \label{Legen2}
\end{eqnarray}  
As follows from Eq.(\ref{Legen2}) that
\begin{eqnarray}
   g_{\alpha\beta,0} \pi^{\alpha\beta} - H_T(x_0, g_{\alpha\beta}, \pi^{\alpha\beta}) \le {\cal L}(x_0, g_{\alpha\beta}, g_{\alpha\beta,0})  \label{Legen3}
\end{eqnarray} 
or, in other words
\begin{eqnarray}
   g_{\alpha\beta,0} \pi^{\alpha\beta} \le H_T(x_0, g_{\alpha\beta}, \pi^{\alpha\beta}) + {\cal L}(x_0, g_{\alpha\beta}, g_{\alpha\beta,0})  \label{Legen3a}
\end{eqnarray} 
The same inequality (Young's inequality) can be derived from Eq.(\ref{Legen1}). 

Now, let us assume that we deal with the actual gravitational field(s) in metirc GR. For the free gravitational field we always have $d-$primary constraints (see above) which are derived from the 
momenta $\pi^{0\alpha}$. This means that we cannot expect that the inequality, Eq.(\ref{Legen3a}) will be obeyed for constrained dynamical systems, e.g., for the free gravitational field. However, 
we can explude such momenta and continue to operate with the space-like components of momenta $\pi^{km}$. As follows from the main text in this case one finds     
\begin{eqnarray}
 H_C(x_0, g_{km}, \pi^{km}) = \max_{g_{km,0}} \Bigl[ g_{km,0} \pi^{km} - {\cal L}(x_0, g_{km}, g_{km,0}) \Bigr] \label{Legen3b}
\end{eqnarray}
The can be re-written in the form
\begin{eqnarray}
   g_{km,0} \pi^{km} \le H_C(x_0, g_{km}, \pi^{km}) + {\cal L}(x_0, g_{km}, g_{km,0})  \label{Legen3c}
\end{eqnarray} 
This inequality must be obeyed for all possible gravitational fields and combinations of such fields. To apply this inequality in actual cases we need either to express the space-like components of 
momenta $\pi^{km}$ in terms of the velocities $g_{km,0}$, or vise versa, the velocities $g_{km,0}$ must be written explicitly in terms of the components of  space-like momenta $\pi^{km}$.

\
\end{document}